\documentclass{PoS}

\def\vt{V773\,Tau}

\def\cut#1 {\sout{#1} }
\def\beq{\begin{equation}}
\def\eneq{\end{equation}}
\def\simgt{\lower.5ex\hbox{$\; \buildrel > \over \sim \;$}}
\def\simlt{\lower.5ex\hbox{$\; \buildrel < \over \sim \;$}}

\title{
Evidence for giant interacting  coronal streamers 
in a pre-main-sequence binary system}

\ShortTitle{Coronal streamers in a pre-main-sequence system}

\author{\speaker{Maria Massi}\\%
        Max Planck f\"ur Radioastronomie \\
        E-mail: \email{mmassi@mpifr-bonn.mpg.de}}

\abstract{
Here we report on  the VLBI discovery   
of solar-like extended streamers  anchored on the  two weak-line T Tauri stars of the
 binary system V773 Tau A. Covering the interbinary distance 
the $\sim$ 20 stellar radii extended 
 streamers  enter in collision during each  stellar rotation with consequent 
occurrence of magnetic reconnection.
Thermal electrons  confined in the streamers become  accelerated to  relativistic speeds
and  emit  synchrotron emission  in the radio band making the magnetic 
streamers "visible" in the VLBI images.
This is different from the solar case  where 
the emission from the  streamers is 
just scattered  photospheric light that would never be observable in distant 
objects.
Evidence of extended solar-like streamers
in T Tauri stars,  thought to be fully
convective, or nearly fully convective objects, 
indicates that the
tachoclinal layer, in this case  either 
not existing  at all
or buried very deeply,
is  not   relevant for the formation
of such solar-like magnetic structures.
}

\FullConference{The 9th European VLBI Network Symposium on The role of VLBI in the Golden Age for Radio Astronomy and EVN Users Meeting\\
		 September 23-26, 2008\\
		 Bologna, Italy}

\begin{document}

%\maketitle

\section{Introduction} \label{introduction}
Coronae are a phenomenon present among a wide variety of objects [1]. 
%(Massi and Preibisch 2006) [1].
Coronae exist not only in solar-like late type stars, but also young  objects, 
not yet in main sequence,   show highly elevated levels of  coronal activity
and even accretion disks, around both compact objects  and protostars,
are postulated as well to have a corona.
From X-ray observations the solar corona results to be formed
by closed bright (i.e. trapping dense and hot plasma) magnetic field lines,  called coronal loops
and open magnetic field lines appearing as dark areas, 
called coronal holes 
%ex[2] %2 (van Driel-Gesztelyi 2006),
through which the fast (400-800 km/sec) solar wind flows [2]. 
%2(Schwenn 2006).
At the top of the closed coronal loops,  
extended streamers are  
observable in visible light during  total solar eclipses,
and  with  satellite-borne coronagraphs as in 
the Solar and Heliospheric Observatory Large Angle
and Spectrometric Coronagraph Experiment (SOHO LASCO).
In the observed radiation, that is the result of the Thomson scattering of 
photospheric light,
one sees a helmet, i.e. a cusp-shaped field line located on top (2 to 4 $R_{\odot}$) of
a closed coronal loop, and a very long streamer which extends
out to $30 R_{\odot}$.
Through the  solar helmet streamers flows the  
slow wind (up to 400 km/s) [2-4].
%(Wang et al. 2007 pag 1346)
%(Schwenn 2006; Suess \& Nerney 2004; Vourlidas 2006;
% Wang et al. 2007). 
During the solar activity cycle the structure of the corona
dramaticaly changes. 
%8(Bravo et al. 1998).  
During sunspot minimum 
high-latitude open field lines form the familiar
polar coronal holes, whereas meridional closed coronal loops form 
around the equator. 
Wind and solar rotation force open magnetic field lines 
from the two hemispheres 
above the closed loops  forming  
the typical
 helmet streamers belt around the equator.
 As the solar activity increases the geometry and orientation of the closed
structures complicates. During the maximum activity closed 
coronal loops  pratically parallel to the equator appear at all latitudes
and helmet streamers  are located everywhere, even
over the Sun's poles [5]. 
%(8 Bravo et al. 1998).
The solar corona itself is therefore already higly structured and dynamic, 
to what extent can we apply our knowledge on the Sun to other coronae?

The theory of dynamo  developed by Parker [6]
%(1979)
 explains how the differential rotation
   generates a toroidal field in  the interior of the sun
in the  transitional layer, called the tachocline,
between  two distinct rotational regimes: the differentially-rotating
 solar convective zone
 and the radiative interior where the rotation
is uniform [7].
%(Miesch 2005).
The corona's building block, i.e. the coronal loop,
is created by the emersion to the surface
by magnetic buoyancy of small portion  of flux tubes
of this  toroidal field.
Very young stellar objects, however, are thought to be 
 nearly fully convective, 
or  fully convective, so the
tachoclinal layer is either buried very deeply,
or does not exist at all.
An important open issue is  therefore 
if different physics of magnetic field generation may determine  large 
differences in  the resulting corona and
if  fully convective objects can have  large scale
solar-like magnetic fields [8].
Here we report on a research based on  radio observations
with the Effelsberg 100-m telescope, 
the Plateau de Bure Interferometer "PdB",
the Very Large Array "VLA" and the Very Long Baseline Array "VLBA". 
The target were two pre-main sequence   objects: 
the two weak-line T Tauri components of the   system V773 Tau A.

\section{The system V773 Tau A}
\label{obs}
\begin{table}
\label{table:log}
\begin{center}
\begin{tabular}{lcc}
\hline
   & Aa(Primary)        & Ab (Secondary) \\
%        & (JD)        &       & (mas)& (mas)       \\
%%    \hline
{\tt Mass$^{[9]}$ (M$\odot$)} & 1.54 & 1.33     \\
{\tt Radius$^{[9]}$ (R$_{\odot}$)} & 2.22 & 1.74    \\
%{\tt Temperature$^a$(K)} & . & .     \\
%%{\tt Age$^a$(10$^6$ yr)} & 3 & 3    \\
{\tt P$_{rot}^{[18]}$}(d) & 2.5 & 1.9    \\
    \hline
    \end{tabular}
    \vspace*{-8pt}
    \end{center}
    %\scriptsize{
    \caption{The binary system V773 Tau A}
    \end{table}
The system V773 Tau (HDE 283447) 
at a  distance  of 136$\pm$5 pc [9],
 in the Taurus-Auriga star-forming region, 
is a quadruplet of young stellar objects within an area  of radius
less than 100 AU (700 mas):
the A component, our target,  is a double-lined spectroscopic binary
of two weak-line T Tauri stars (Aa and Ab),
the B and C components, with   
significant
infrared excess, are at an earlier 
evolutionary stage.
The two components V773 Tau  Aa and Ab (Table 1), 
with  orbital period  of  51.1033 d and  
orbital separation (2$a$)  of 5.5. mas (0.8 AU),
are still in the quasi-isothermal vertical track  (contraction phase) 
and have  an  estimated age of 3 $\pm$ 1 Myr [9].

\section{Interbinary collisions}
V773 Tau A  was observed with the Effelsberg 100-m telescope 
over a frequency range spanning from 8 GHz (3.6 cm) to 40 GHz (7 mm)
for  522 days during which where collected 110 samples.
The source was also observed with the VLA for
one month at a frequency of 14.96 GHz (2.0 cm). 
Figure 2 in Massi, Menten and Neidh\"ofer [10], presents  
the data  folded with the orbital  period ( 51~d) and reveals 
that the system exhibits persistent flaring activity that
 gradually increases from few mJy  at apoastron
to very  strong ($S\sim 100$ mJy) flares around  periastron.

There exists  a close relationship between flares and interaction of magnetic structures.
In fact, as  observed on the Sun, flares can be  triggered by
interactions between new and older emergences of magnetic flux
in the same  area [11] 
i.e. when in its emergence, because of magnetic buoyancy,
on the surface, one coronal loop
happens to  intrude into other already established coronal loop,  magnetic reconnection occurs.
Part  
of the 
magnetic energy released during reconnection goes to accelerate 
a fraction of the thermal electrons  trapped in the flaring loop
and a  population of relativistic electrons  is 
produced [12].
%(Drake et al. 2004). 
%http://www.newton.ac.uk/preprints/NI04028.pdf
These   relativistic particles
gyrating around the magnetic field lines of the coronal loop,
where they are confined, generate  synchrotron emission in the radio band. 
%(Parkes 1979; Bastian et al. 1998).
Applying this knowledge of solar flares to V773 Tau A,
it is clear that in this system the observed  relationship  between  intensity of the flare
occurrence and distance of the two stars
indicates another, new mechanism of magnetic interaction, that of interacting coronae
 and therefore with magnetic reconnection  
taking place far out from the stellar surfaces, where the two coronae interact 
with each other.
Knowing that the distance between the two stars at apoastron and at periastron is $52 R_*$ and $30 R_*$, respectively,
interbinary collisions would then imply very large coronal structures  with
a size  of $15 R_* \le H \leq 26 R_*$.
As a matter of fact  a large magnetic structure  
has been imaged by
using Very Long Baseline Interferometry  [13]
showing a large  halo ($\ge$ 3 mas, $\sim$40 $R_*$)
 surrounding  two compact components 
 separated 1.14 mas ($\sim$16 $R_*$).
On other hand Skinner et al. [14] interpreted the light curve of a hard X-ray flare in
V773 Tau A 
as being due to the rotational modulation of the emitting flaring region,
determining  a size of $H\le 0.6\, R_*$.
Tsuboi et al. [15] interpreted the decay of another hard X-ray flare
as being due to radiative cooling,   obtaining  a  size
of $1.4\, R_*$.
Therefore,   radio and X-ray emission clearly come from spatially separated regions:
 smaller
ones those associated with X-ray emission and
larger  ones those associated with radio emission.
Further confirmation of the presence of two structures comes from
the multiwavelength campaign on V773 Tau A
carried out by Feigelson et al. [16]
showing    radio variability combined with
a  steady X-ray flux.
Following the solar analogy we postulated solar-like  extended    
streamers    
located at the top of the observed stellar sized X-ray emitting coronal loop.
V773 Ta A was observed with the IRAM Plateau de Bure Interferometer
at  3 mm and 
a complete (onset and decay) 
flare of  360 $\pm$ 17 mJy  was observed close to periastron ($\Phi=0.1$).
The  flare decay (e-folding time of 2.31$\pm$ 0.19 hours ) was  
 modeled with  a slow leakage of relativistic  electrons trapped in a
magnetic structure with  two mirror points: 
one close to the star (at  2-5 R*  ) i.e. the helmet 
and the other 
to the top of the streamer (at  10- 20 R* ) 
 i.e. where the two streamers interact with each other [17].
%(Massi et al. 2006).
At the distance of 136 pc [9], one stellar radius ($R_*\simeq 2 R_{\odot}$),
corresponds to 0.069 mas.
Only the high resolution of VLBI/VLBA can probe the proposed model.
\section{VLBA observations}
\subsection{Interbinary collision of streamers catch in the act}
We observed \vt~ A for seven consecutive days (listed as A to G in Table 2) 
at 8.4\,GHz with  
the VLBA together 
with the  Effelsberg 100-m radio telescope (VLBA+EB) [18].
\begin{table}
\label{table:log}
\begin{center}
\begin{tabular}{ccccc}
\hline
Run     & Date        & $\Phi$& Ha-Hb&Aa-Ab\\
        & (JD)        &       & (mas)& (mas)       \\
    \hline
{\tt A} & 2453076.34 & 0.32 & &2.7    \\
{\tt B} & 2453077.34 & 0.34 &2.8&2.9    \\
{\tt C} & 2453078.34 & 0.35 &3.2&3.1   \\
{\tt D} & 2453079.34 & 0.37 &3.2&3.2   \\
{\tt E} & 2453080.34 & 0.39 &3.4&3.3  \\
{\tt F} & 2453081.34 & 0.41 &3.5&3.4  \\
{\tt G} & 2453082.34 & 0.43 &3.1&3.5  \\
    \hline
    \end{tabular}
    \vspace*{-8pt}
    \end{center}
    %\scriptsize{
    \caption{
    Stellar and helmet separations in the young stellar binary system V773~Tau~A during each VLBA+EB observation. Julian day 2453076.34 corresponds to  March 11, 2004 at 20:10.
Ha and Hb are the two radio peaks in the VLBI images
associated with the helmets.
Aa and Ab are the primary  (V773 Tau Aa) and secondary (V773 Tau Ab)
stars  of the system respectively.
The error associated to the  separations
is  $\pm 0.1$mas.
    At  a distance  of 136 pc,
    one stellar radius, $R_*\simeq 2 R_{\odot}$,  corresponds to 0.069 mas.
    }
    \end{table}
Whereas in  the image of the first run
there is only one 3$\sigma$ feature,
two structures  are  present 
 in all the other images (see here Fig. ~\ref{vlbi}-B,E,G),
here called North-East, NE, and South-West, SW. The two structures
can be quite complex as in Fig. \ref{vlbi}-B where they show two components
each  (``a", ``b" for the NE and ``c", ``d" for the SW).
 The location  of the stars in the radio images of Fig. ~\ref{vlbi},
is given as a  cross [18].
In Fig. ~\ref{vlbi}-B,
 the two features are  extended  18$R_*$   each, which are the expected $H$ size
for the postulated streemers.
The relativistic particles,
produced by magnetic reconnection at the top of the streamer,
spiralling around magnetic field lines emit synchrotron
radiation and make the whole streamers "visible"  in the radio band.

Numerical calculations based on the general equation of radiative
 transfer show that the distributions of energetic electrons 
along a solar flare coonal loop are highly inhomogeneous: 
accelerated electrons are concentrated 
where the magnetic field is stronger [19].
Radio images, during  solar  flares show  
brightness peaks  located 
at the footpoints of the coronal loop [20], which are the mirror points of the trapping structure.
From that it follows that during a flare  
the brightness   peaks in the radio images reveal the position of
the mirror points.
Whereas in the solar coronal loop case   the mirror points are the footpoints
of the loop, in V773 Tau A  the two mirror points for each of the two NE and SW features,
are one close to the star and the other displaced several stellar radii in the upper corona.
The  model  based on the flare decay of the flare at 3 mm, 
derived a first mirror point  
close to the star (at  2-5 R*  ) i.e. the helmet.
The  peaks "a" and "c", that   are the two closest peaks to the two related  stars,
are therefore the two helmets.
The second 
mirror point in the model is located    
at the top of the streamer (at  10- 20 R* )  i.e. where the two streamers interact with each other.
As a matter of fact the two extremes "b" and "d" of the two features in Fig. 
\ref{vlbi}-B
not only are  at the predicted distance from their respective
 helmets, but  also are close each other, as one expect from interacting streamers. 
It seems that in run B  (see sketch of the system in Fig. \ref{vlbi})
 we were lucky enough to catch the interaction of the two streamers in the act.
\subsection{Streamers anchored on rotating stars}

\begin{figure*}[]
\begin{center}
\includegraphics[width=.33\textheight]{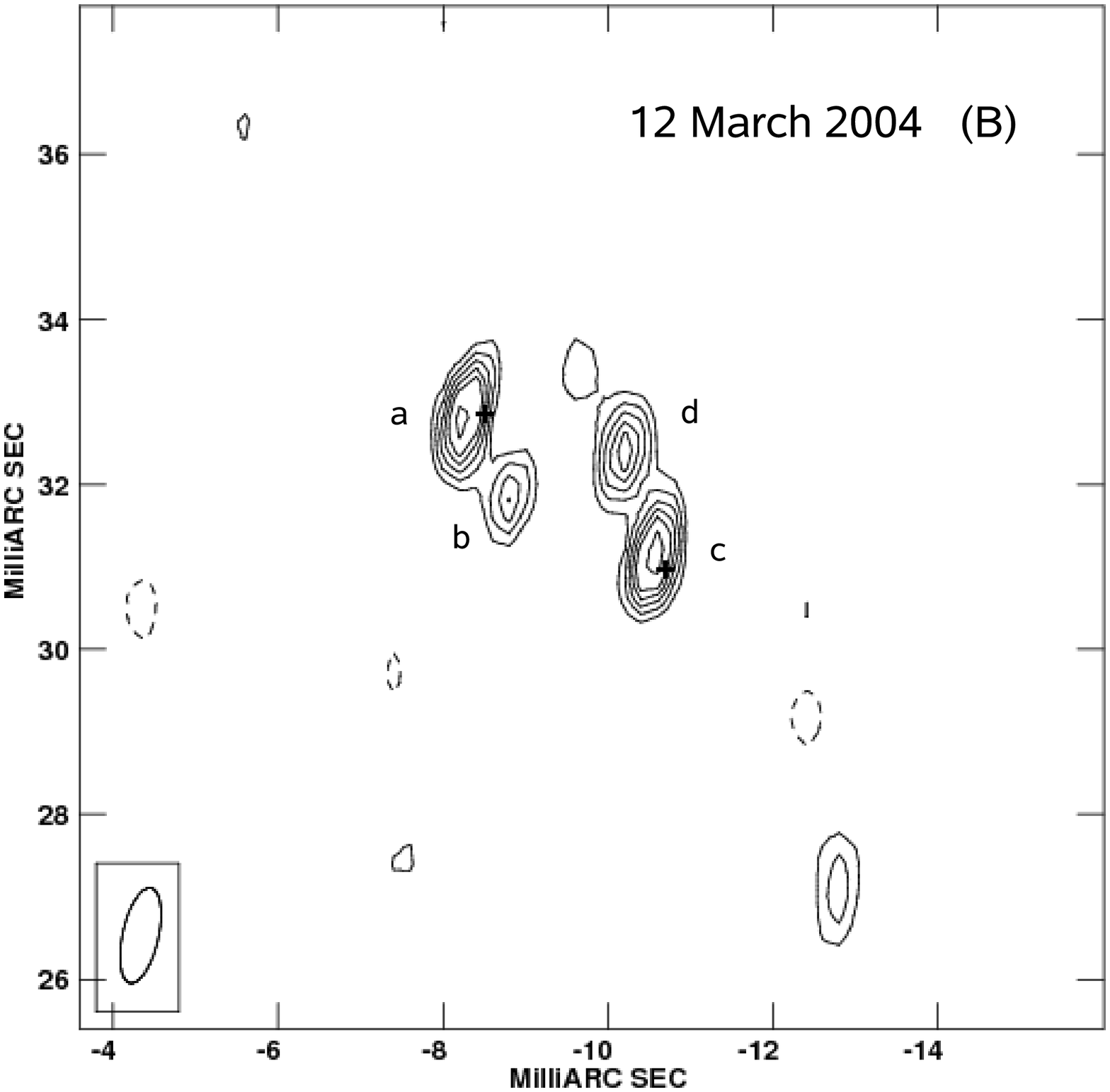}
\includegraphics[width=.33\textheight]{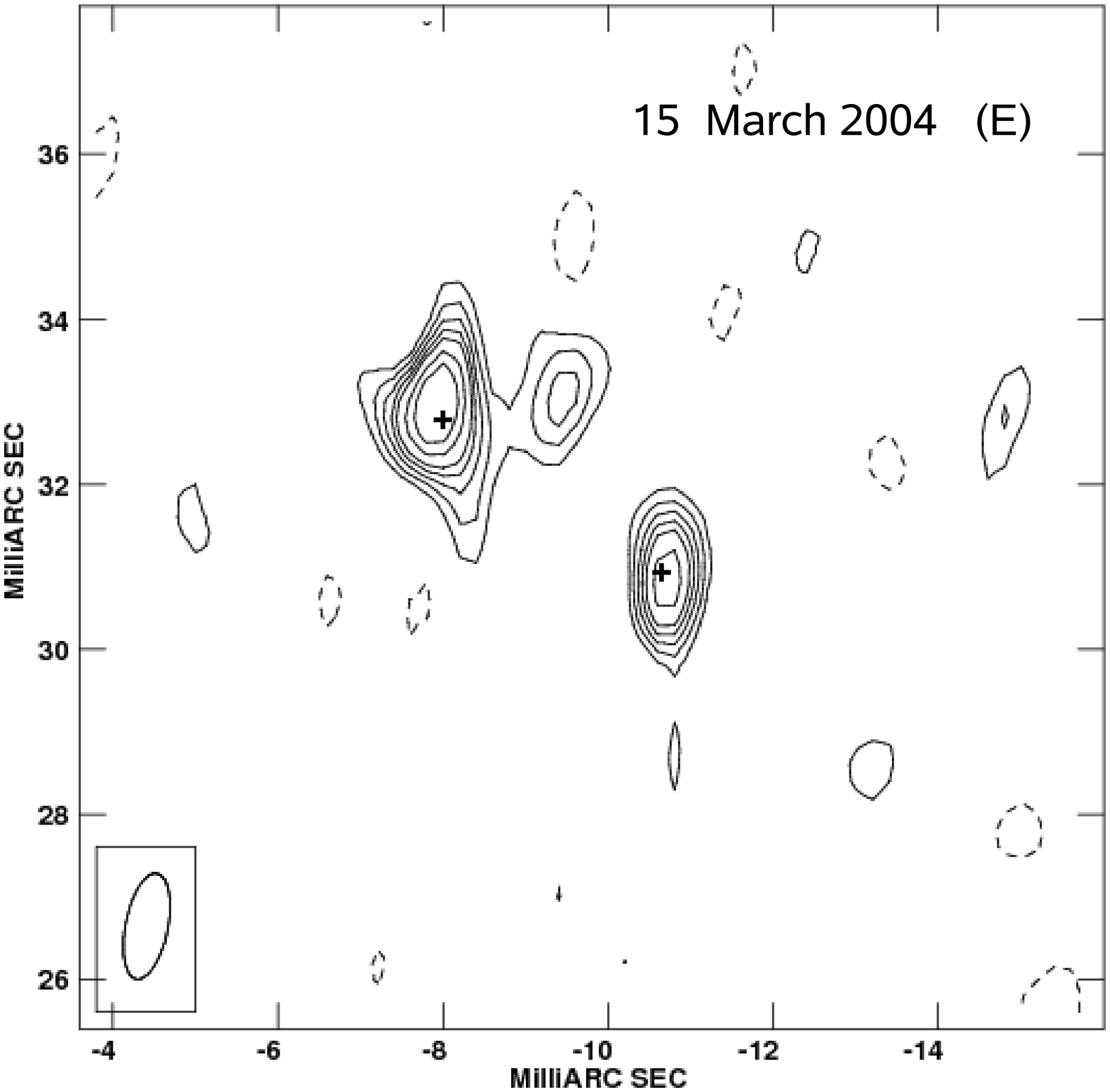}
\includegraphics[width=.33\textheight]{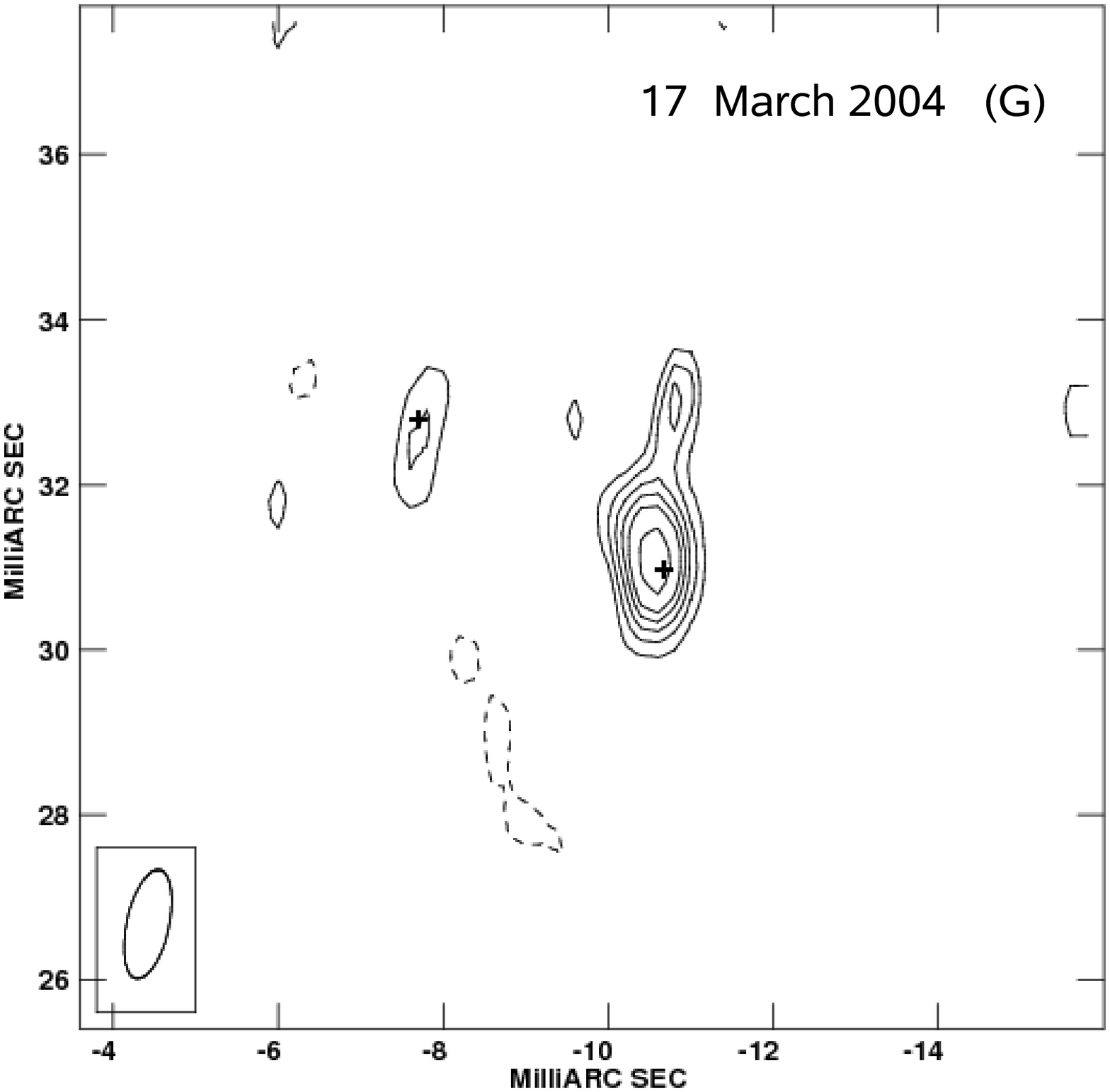}
\includegraphics[width=.33\textheight]{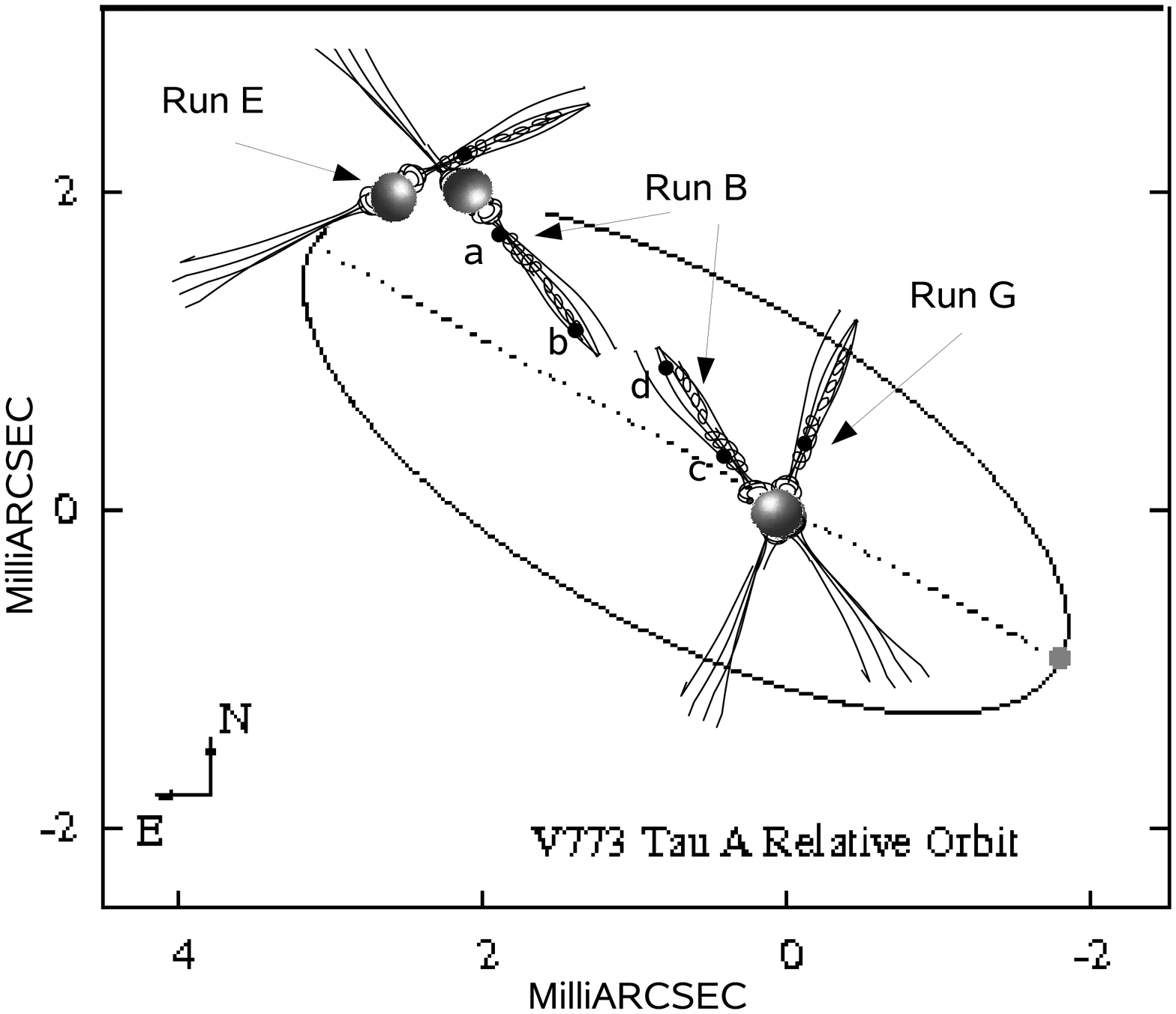}
\end{center}
 \caption{
% Seven consecutive 
8.4\,GHz VLBA+EB images of the young stellar binary system V773~Tau A.
Stellar positions are indicated by crosses [18].
  The peak flux density in the three B, E and G images
  are  0.5,  0.7, and  0.5  mJy/beam respectively.
  Contour levels are $-$1, 1, 1.5, 2, 2.5, 3, 4 and 5 times the
  3 $\sigma$ (0.1 mJy/beam).
  The beam size is shown in the bottom left corner of each panel ($\sim$1.3$\times$0.5 mas).
 The  sketch of the system   shows 
 helmet streamers
 anchored on rotating stars (not to scale).
}
\label{vlbi}
\end{figure*}

The 100-m telescope monitoring shows (Fig. 3 in [18]) that at the time of  image
Fig. 1-E the flare  is already in its decaying phase.
   In fact,
   while a total flux density of 12 mJy was measured with the Effelsberg 100-m telescope
   two hours before
   the  VLBA+EB  observation,
   only 6 mJy were observed when the observation begun.
If the collision occurs when the streamers are aligned as in Fig. 1-B,
one expects that few hours ($>$ 2 h) after the collision,
because of the short rotation period ($\simeq 2$ d, Table 1)
the streamer, anchored on the rotating star,
should appear in the image appreciably rotated with respect to 
the position angle of the streamer 
in Fig. 1-B (PA=208$^{\circ}$).
Indeed the position angle of the NW feature of Fig.~\ref{vlbi}-E 
is of PA=273$^{\circ}$.
Besides the rotation, another fact  indicates that time has elapsed from the
streamer-streamer collision:
the   absence of the upper mirror point in the SW feature.
The solar streamers are open structures. Therefore,  one expects that the compressed 
magnetic field lines at the top  of the streamer ("d" and "b")
relax back 
to an open configuration, i.e.,  the peaks at "d" and "b"  should have
a faster decay with respect to the helmets "a" and "c".
The relativistic
electrons diffuse from the relaxing upper mirror points  
in progressively larger structures:
 in Fig.~\ref{vlbi}-B, that is just after/during
a collision, the sources are point-like, whereas
in   Fig.~\ref{vlbi}-E,
showing an already evolved situation, the sources are extended.
With this respect, in the last observation of March 17
(Fig. 1-G), the facts that on one hand the helmet streamer anchored on the
primary
star appears substantially rotated and extended
and  on the other hand, there is a lack of the upper mirror in the NE
structure, suggest again a flare decay scenario as  for Fig. 1-E, only applied to the other star.
\section{Conclusions and Discussion }
In our VLBA observations   individual streamers 
can be distingueshed in each of the two stars of the system V773 Tau A.
The complete extent of the helmet streamers
is  the observed value of   18 $R_*$, between helmet and upper mirror point, plus the distance,
of a  few stellar radii,
      of the helmet itself from stellar surface.
% (a $\sim$ 6 $R_*$)[19].
The practically permanent flaring activity in V773 Tau A,
from a level of a few mJy  around apoastron (52 $R_*$)
to more than  100 mJy  at periastron (30 $R_*$) [10]
is explained by the interaction of these 
extended streamers.
At the short periastron  distance  the two coronae nearly   
 overlap, giving rise to  the
observed giant flares (i.e. up to 360 mJy at $\lambda$ 3mm [17]) 
and to the large structure imaged by VLBI by Phillips and collaborators [13].

Our observation [18] proves  that solar-like streamers of more than 20 $R_*$ 
exist  also in fully / nearly fully convective objects, that is in objects where
the tachocline is  missing or 
buried very deeply.
Besides, the helmet streamers are an important magnetic configuration for themselves.
Through the  solar helmet streamers flows the  
slow wind (up to 400 km/s).
The plasma flow  
presents  density fluctuations, i.e. "blobs" or "plasmoids" 
that seem to be the product of reconnection or small-scale eruptions
at the cusp of the helmet streamer.
On other hand   helmet streamers are related with 
coronal mass ejections, where   plasma clouds leave the sun
in violent way (impulsive CMEs) ([2] and references therein).
%and may reach 2000 km/s (Schwenn p. 64).
Because of these characteristics of being associated to plasma motions/ejections
and because of their topology, 
i.e.  open field lines pushed above a closed loop,
the helmet streamers are invoked as a  key magnetic feature  bridging
the gap between  corona  and jet.
In fact, they have been proposed 
in the process of jet formation 
in  microquasars/AGN [21]
and also  in young stellar objects [22,23].
The  discovery  of helmet streamers
in stars other than the Sun, at a   
wavelength observable at high resolution, could therefore lead to  
a deeper understanding of these important magnetic structures and their possible related processes.
\acknowledgments
Based on observations with the  Effelsberg 100-m
telescope of the
Max-Planck-Institut f\"ur Radioastronomie (MPIfR) and
the Very Long Baseline Array (VLBA).
The   VLBA is a   facility of the National Radio Astronomy Observatory (NRAO),
operated by
Associated Universities Inc. under a cooperative agreement with the National Science Foundation
(NSF).

\end{document}